# Experimental Investigation of Thickness Effects on Mixed-mode I/II Fracture of an Aluminum Alloy

Wanlin Guo[1, 2*], Huiru Dong[2, 3], Zheng Yang[2]

[1]*Institute of Nanoscience, Nanjing University of Aeronautics and Astronautics, Nanjing 210016, China*

[2] *Department of Mechanics, Xi'an Jiaotong University, Xi'an 710049, China*

[3] *The engineering institute, Air Force Engineering University, Xi'an 710038, China*

**Abstract:** Experimental investigations on the effect of thicknesses on mixed-mode I/II fracture are performed with an aluminum alloy with thicknesses of 2, 4, 8 and 14 mm. It is found that under pure-mode I loading condition, the loading capacity per unit thickness monotonically decreases with increasing thickness. However, under mixed-mode loading condition, the dependence of loading capacity on the thickness becomes very weak. Although the measured crack initiation angles from the mid-plane of the fractured specimens coincide well with typical existed criteria in the thin and thick specimens under all the mixed-mode conditions except pure-mode II, exceptional low crack initiation angles are found in the 8mm-thick specimens. Although ductile dimples are the dominated microscopic features in all the specimens, the size of the final voids varies significantly with specimen thickness and loading mode. The ductile dimples in the crack initiation zone become larger and shallower as the specimen thickness decreasing.

---

[*]Corresponding author: wlguo@nuaa.edu.cn, Fax: +86 25 84895827

# 1. Introduction

Fracture mechanics has long been playing an important role in integrity assessment of engineering structures, and experimental data of materials obtained by using standard or simple specimens in laboratory have been proven to be the necessary fundament. However, the loading condition, stress states and geometry of the specimens in laboratory are far from that of the structural components in engineering. Most of the cracks in engineering structures usually present three-dimensional (3D) mixed-mode in spite of lots of studies focused on the single mode crack. In the last decade, 3D fracture theory [1-8] along with constraint theories [9-14] were developed to explain the discrepancy in fracture toughness caused by changes in geometry and stress state. As the most typical 3D problem, the effect of thickness on fracture has been intensively studied theoretically, numerically and experimentally [15-18]. It is shown that thickness has strong influence on the crack-tip fields as well as the fracture toughness of materials. However, most of the 3D studies aforementioned are confined to pure mode I crack. In fact, crack tunneling, curving and slanting are inevitable during crack growth even in an initially pure mode I cracked body [19,20]. Therefore, 3D mixed-mode fracture is much more important in engineering and need extensive study.

Since the maximum tangential stress (MTS) criterion was proposed by Erdogan and Sih [21] for mixed-mode crack initiation, mixed-mode fracture has long been studied and many mixed-mode fracture criteria have been put forward including the minimum strain energy density criterion [22], the maximum stress triaxiality criterion [23], and the crack-tip opening displacement (CTOD) based fracture criterion [24]. However, the limited available experimental data exhibits large scatter, and does not support a certain criterion for different loading conditions [25, 26]. The most important potential reason may be the influences of specimen thickness and geometry used in different experiments. It is deserving that the constraint effect on mixed-mode fracture has been taken into account in some studies [27-31]. Recently Khan and Khraisheh [32] proposed a new criterion for mixed-mode fracture initiation based on the crack-tip plastic core region. The proposed criterion states that the crack extends in the direction of the local or global minimum of the plastic core region boundary depending on the resultant stress state at the crack-tip. We know that the stress state near the crack-tip depends on the thickness of the cracked plate, so it can be expected that there may be some relations between the thickness of specimen and crack initiation under mixed-mode loading conditions. However, to the best knowledge of the authors, no systematical experimental investigation has been conducted on the simplest 3D effect on any materials up to now.

For a better understanding of mixed-mode fracture, not only the corresponding 3D or constraint theories are highly needed, but also the most fundamental 3D phenomena have to be elucidated experimentally. The main aim of the present work is to investigate experimentally the effect of thickness on mixed-mode fracture of materials. Systematical experimental investigations are performed on mixed-mode fracture with compact-tension-shear (CTS) specimens of a LC4-CS aluminum alloy with combinations of four thicknesses from 2 to 4 mm and seven loading angles from pure mode I to pure mode II being considered. Exceptional thickness effects on some important features such as the load versus crack opening displacement curve, macroscopic and microscopic fractographs as well as the crack initiation direction are revealed the first time. Some existing criterions for mixed-mode crack initiation are discussed against the background of the test data and the problems needed further study are proposed.

## 2. Experiment

### 2.1. Material and specimens

The material used was a high-strength aircraft structural aluminum alloy LC4-CS. The typical chemical composition for the alloy is shown in Table 1. The yield stress and ultimate tensile stress of the material are 485 MPa and 510 MPa, respectively.

The CTS specimens were chosen for all fracture testing and fabricated in the T-L orientation. The specimens with four thicknesses of 2, 4, 8, and 14 mm are designed to reflect the variation in three-dimensional constraint near the crack-tip. The same in-plane geometry is used to all specimens to avoid change in the in-plane constraint. All of the specimens were machined from a single 14 mm thick mother plate to avoid lot-to-lot variation. The microstructure of the material was found to be homogeneous along the thickness. For the 14 mm thick specimens, only final surface machining was required in the thickness direction. To produce specimens with thickness less than 14 mm, the outer surfaces of the 14 mm plate were machined by mechanical milling to reduce the plate to the desired thickness. During this thickness machining process, care was taken to alternate the material removal side of the plate to minimize possible out-of-plane warping. An initial 2 mm wide V-rooted notch in each CTS specimen was mechanically cut to depth of $a_n$=25 mm, with notch depth to specimen width ratio of $a_n/w$=0.5.

Before the fracture tests, the notched specimens were fatigue precracked under constant amplitude loading with a stress ratio slightly above zero. The final total length of the notch and the fatigue crack, or the initial crack length is about $a_0$~27 mm ($a_0/w$ is about 0.54). The detailed in-plane geometry of the CTS specimen is shown in Fig.1. In every specimen, the fatigue crack had a tunneling curved front and the initial crack length $a_0$ is taken as the average along the thickness in the way recommended by ASTM standard E399-83. As the fatigue precracking process was controlled by examining the fatigue crack length from the surfaces of the specimens, slight larger average $a_0/w$ is obtained for thicker specimens for the tunneling crack front in the interior.

## 2.2. Experimental fixture and procedures

The special loading fixture designed by Richard [33], shown in Fig.2, was employed. The loading angle $\phi$ is defined to be the angle between the initial crack direction and the direction of the applied loading. The appropriate selection of the loading holes in the loading fixture results in an entire range of mode I/II loading from pure mode I ($\phi$=90°) to pure mode II ($\phi$=0°).

All experiments were performed under laboratory condition with temperature of 17°C and humidity of 52% RH. Each specimen was loaded in a displacement–controlled 100 kN MTS 810 testing machine at a constant cross-head rate of 0.4 mm/min.

During the test, a COD clip-gauge was mounted across the notch at one side of the specimen for measurement of the crack mouth opening displacement (COD). Traces of load versus COD were recorded automatically by the test system every 0.05 mm of displacement.

A zoom-field microscope, a CCD camera, and commercial hardware/software for facilitating the image monitoring and acquisition were used for crack initiation measurements. The CCD camera output was routed through the digital image acquisition system and into a computer for monitoring and storage. Figure 3 shows the crack-tip appearances of a 2 mm thick specimen under 60° loading angle during the fracture test process. Figure 3(a) shows the original precracked crack with initial length of $a_0$. With increasing load, the crack becomes wider. When the load reaches a critical value, the original crack will initiate to grow, as shown in Fig.3(b). Certain crack-tip opening and slipping displacements can be observed, and a small crack-tip plastically deformed zone (the shiny zone) appears in front of the crack-tip in certain angle to the initial crack plane. As shown by Fig.3(c), near the maximum load level, notable crack growth can be observed and the crack-tip opening and slipping displacements become significant and the plastically deformed zone becomes larger. The confined plastic zone ahead of the crack-tip and plastic wake left by the crack growth show that the ductility of the tested material is relatively week even in the thinnest specimens.

After fracture, all specimens were sectioned into two halves at the mid-thickness by the electric spark wire-cutting method. Crack initiation directions were measured with a universal bevel protractor at the middle plane as crack initiation always occurs first at the center of the specimen. Measurement at the middle plane can also avoid the influence of the plastic shear lips formed in the crack wake near the free surfaces of the specimens. Fracture surfaces of the fractured specimens were examined by means of macroscopic observation and a scanning electron microscope HITACHI S-2700.

## 3. Experimental results and analyses

3.1. Load-crack opening displacement curves

*3.1.1. Effect of thickness on the load-crack opening displacement curves*

The load-crack opening displacement (load-COD) curves of various thicknesses under each mixed-mode loading condition are drawn together to find out the thickness effect on the load-COD curves. Because the in-plane geometry of the specimens remains unchanged except the slight difference in $a_0$, the loading capacity of a specimen is represented by the load normalized by thickness, as shown in Fig.4. Figure 4(a) shows the load-COD curves of various thicknesses under pure mode I loading condition ($\phi = 90°$). As can be seen clearly, the maximum load-carrying capacity per unit thickness decreases significantly with increasing specimen thickness. The capacity of the 14 mm thick specimen is only 59% of that of the 2 mm specimen.

Figures 4(b)~4(d) present the results obtained with the loading angles $\phi = 60°$, $45°$ and $30°$, respectively. It is interesting to find that the loading capacity per thickness of the 4 mm thick specimens is the maximum in most mixed-mode loading conditions, and the capacity of the 2 mm, 8 mm and 14 mm specimens are almost the same except of the 14 mm curve in Fig.4(e). We check the data and find that the fatigue precrack length of this 14 mm specimen ($a_0/w$=0.570) is obviously larger than that of the rest specimens ($a_0/w$=0.535, 0.531 and 0.524 for the 2, 4 and 8 mm specimens respectively). The reason why $P_{max}/B$ of the 2 mm specimens is not the maximum may is partially due to the out-of-plane buckling, especially for smaller loading angle $\phi$, as the initial crack length of all the 2mm specimens is nearly the same or slightly smaller than that of the rest sets of specimens.

The situation for $\phi=75°$ is quite similar to the cases of $\phi=30°$ to $60°$, with the load capacity per thickness is highest at 4 mm (2.2 kN/mm) and the capacity for 2, 8 and 14 mm specimens is in range of 1.5 to 2.0 kN/mm and does not change monotonically with thickness. This means that the monotonic decrease of load capacity in pure mode I condition can only occur hold good in situations of very small fraction of mode II compoment.

As can be seen in Fig.4(e), the influence of thickness on the loading capacity becomes even weaker at $\phi=15°$. The slight difference in loading capacity is due to the slight difference in the initial crack length and uncertainties in the experiment. Under pure mode II loading, it has been shown by detailed 3D finite element analyses that the crack-tip stress and strain fields, as well as the stress state, do not change along the thickness [34]. As can be expected, with the loading angle decreasing from 90°, the mode II component will increase and the thickness effects will decrease and finally disappear at $\phi=0°$. As the COD gauge cannot be fixed well in the pure mode II test, no complete load-COD curve is recorded and we present the maximum loading capacity in this case in Fig.5 against the thickness. It can be found that the test data are nearly independent of thickness in overall.

To clarify the influence of the slight difference in the initial crack length, we represent the load-COD curves for $\phi=90°$ to 15° in Fig.6 by use of the effective stress intensity factor evaluated at the initial crack length $a_0$

$$K_{eff} = \sqrt{K_I^2 + K_{II}^2} \ . \tag{1}$$

Where $K_I$ and $K_{II}$ are the components of the stress intensity factor for mode I and mode II loading as shown latter by equation (4) following the 2D theory. It should be pointed out that such a $K_{eff}$ is not the real stress intensity factor at given COD as no crack growth is taken into account and no 3D distribution is considered. We can simply use the $K_{eff}$ as an alternative measure of the normalized load level, which can partially eliminate the influence of the slightly difference in initial crack length. Comparison of Fig.6 and Fig.4 shows that the influence of initial crack length does not change the general trend. An exceptional case is the 14 mm thick specimen in Fig.6(c). This specimen has initial crack length of $a_0/w$=0.573, much higher the crack length in the rest cases in Fig.6(c), with $a_0/W$≤0.534 and this difference leads to a change in the $K_{eff}$ as high as 18%. It is important that Fig.6 confirms again that even at $\phi=75^o$, with $K_{II}$ of about 12% $K_I$, the strong thickness dependent loading capacity at pure mode I has nearly disappeared completely. Therefore, the thickness effects on the fracture of mixed-mode I/II cracked specimens cannot be estimated by some simple superposition of the effects in mode I and mode II conditions.

*3.1.2 Effect of loading angle on the load-crack opening displacement curves*

For each thickness, the load-COD curves under different loading angles are compared in Fig.7. For the 14 mm specimens as shown in Fig.6(a), when loading angle $\phi$ increases from 0 to 45$^o$ the load capacity decreases monotonously. However, when $\phi$ changes between 45$^o$ to 75$^o$, the load-COD curves are nearly the same. When $\phi$ increases further to 90$^o$, the load capacity goes down remarkably.

For the 4 mm and 8 mm specimens, the overall trend shown in Figs.6(b) and (c) is that the loading capacity decreases with increasing $\phi$, although a few curves do not obey the general trend due to the slight difference in initial crack length and other uncertainties in the experiments.

In the 2 mm specimens, things become much more complicated mainly because of the buckling and out-of-plane twisting whose degree depends on the loading angle. For $\phi \leq 60^o$, the overall trend is similar as that of the 4 and 8 mm specimens, or the loading capacity decreases with increasing $\phi$. However, when $\phi \geq 60^o$ where the mode I loading becomes more dominated and the out-of-plane buckling and twisting may become some less possible with increasing $\phi$, a reverse trend is observed. This reverse trend cannot be explained by the difference in the initial crack length as all the 2 mm specimens tested with $\phi \geq 60^\circ$ is quite stable, with $0.527 \leq a_0/W \leq 0.529$. Consequently, the dependence of the loading capacity on the mixed-mode loading condition is dependent upon the thickness of specimen.

**3.2. Macroscopic fracture surfaces**

Different fracture mechanism may be excitated by variations in loading conditions and stress states. So analyses of the morphology of the fracture surfaces may provide better understandings of the complexity of mixed-mode fracture of finite thickness specimens. As illustrated in Fig.8(a), a typical fracture surface under mode I loading condition consists of four distinct regions: fatigue pre-cracking, flat fracture, flat-to-slant transition, and slant fracture [18, 35]. However, there is wide variability in the macroscopic fractographs for the mixed-mode fracture of various thickness specimens in this study. As shown in Fig.8(b), the fracture surface for the 14 mm thick specimen shows a flat ductile fracture in the center of the plate, whereas the shear lips dominate in the vicinity of the free surfaces except under pure mode II loading condition, the fractions of flat fracture and shear lip vary with the loading angle. The shear lips become smaller and irregular as the mode II fraction increases, and almost disappear when the loading angle decreases to $\phi = 15°$. Flat-to-slant transition and slant fracture do not appear on the fracture surface when $\phi \leq 15°$.

For the thin 2 mm and 4 mm specimens, the profiles of the mixed-mode fracture surfaces are similar as shown schematically in Fig.8(c). The fracture surface is composed of a small triangular flat-to-slant transition zone and a long slant fracture zone, and no through-thickness flat fracture zone exists. It can be seen from Figs.8(b) and 8(c) that the crack kinks from the original crack plane under mixed-mode loading conditions. The fracture profile of the 8 mm thick specimen is more complex, comprising flat fracture with shear lips and slant fracture. The flat fracture zone is so small that the fracture surface adopts a V shape in the 8 mm thick specimen. The fracture under pure mode II loading condition for any thickness is pure in-plane shear, neither shear slip nor slant fracture can be observed.

The above results indicate that the macroscopic fracture surface morphology of normal fracture under pure mode I and mixed-mode loading condition varies obviously with specimen thickness and loading angle. Shear lips that usually being believed to reflect the size of the crack-tip plastic zone with low out-of-plane constraint provide important information concerning the stress state in the specimen. For thin specimens (such as $B$=2 and 4 mm), the size of the plastic zone is large and failure occurs on the plane containing the maximum shear stress, so slant fracture forms. In addition, for thick specimens namely when the plane strain state dominates, the plastic zone is much smaller and the fracture surface is oriented nearly at 90° to the loading direction. Since plane stress prevails at the free surfaces of a specimen, shear lips are always present in mixed-mode fracture, albeit small in some instances. The flat-to-slant transition is really a highly three-dimensional process. The 8 mm thick specimen ranks between thin and thick, so the macroscopic fracture is more complex. Under pure mode II loading condition, all specimens fracture in an in-plane shear mode and thickness has no effect on the macroscopic fracture surface morphology.

**3.3 Microscopic fractographs**

Figures 9(a)-(c) shows the SEM fractographs of the broken specimens in the crack initiation region under pure mode I loading condition corresponding to $B$=14, 8 and 4 mm, respectively. Many dimples are observed on the fracture surfaces. As thickness decreases, the dimples stretch larger and shallower. Fractured inclusions are observed in Fig.9(b) (see arrow). Figure 10 shows the SEM fractographs of the 14 mm and 8 mm thick specimens under pure mode II loading condition. Many closely spaced small voids are observed. Some ripple marks exist in the 14 mm thick specimen shown in Fig.10(a). In Figs.11(a)-11(e), fractographs of the specimens under 45° loading condition are presented for thickness $B$=14, 8, 4 and 2 mm, respectively. On the fracture surfaces of the 14 mm thick specimens, small ductile dimples are the dominated features, while large ductile voids are the main features on the fracture surfaces of the 4 and 2 mm thick specimens. The 8 mm thick specimen has a mixture of both small dimples and large ductile voids, with some large ductile ridges. It is very interesting that the shear-type voids shown in Fig. 11(d) and tensile-type voids shown in Fig. 11(e) coexist in the 2 mm thick specimen. In general, figures 9-11 show that voids become larger in thinner specimens for all the mode I, mode II and mixed-mode I/II loading conditions.

The fractographic characterizations demonstrate that the fracture of LC4-CS

aluminum alloy is ductile, i.e. consists of microvoid nucleation, growth and coalescence. Thickness has significant effect on the microscopic fractographs of mixed-mode fracture. In thicker specimens, the ductile voids cannot grow larger and the ductility is relatively lower. As thickness decreases, both tensile-type dimples and shear-type voids stretch larger and shallower with ductile deformation can be better developed. Consequently, both the loading angle and specimen thickness have significant effects on the fracture mechanism.

### 3.4. Crack initiation angles

Two dimensional mixed-mode fracture investigations assume that from the practical point of view, the elastic solutions are sufficient for predicting the crack propagating direction, even for very ductile materials [36, 37]. As shown by the confined plastic zone in Fig.3, small scale yielding assumption can be adopted for the present tests. In the following sections, the crack initiation angle will be predicted by some respective approaches based upon the elastic solutions and validated against our experimental results. As shown in Fig.12, the crack initiation angle, $\theta_0$, for the mixed-mode crack is defined on the mid-plane of the specimen as the angle between the original crack plane and the tangent to the propagation trace within the very beginning stage of crack growth. It is observed from the fracture profiles that the crack propagation trace is nearly a straight line at the first 2-3 mm crack extension, within the flat center region. Thus, the practical measurement of the mixed-mode crack initiation angle, $\theta_0$, is made by the straight fitting tangent line along the crack trace starting from the initiation point and within 3 mm. Crack initiation angles of all specimens are obtained by this method.

Figure 13 shows the variation of the average experimental results of crack initiation angles versus thickness under various loading conditions. It indicates that the crack initiation direction for all thicknesses under pure mode I loading is along the original crack direction. All cracks kink under mixed-mode loadings. $\theta_0$ increases with increasing mode II loading fraction or decreasing loading angle except for the pure mode II loading condition. The crack initiation direction in the 2 mm thick specimens is close to that in the 14 mm thick specimens. However, it is slightly higher in the 4 mm thick specimens and significantly lower in the 8 mm thick specimens. Such an exceptional thickness dependent mixed-mode crack initiation angle has not reported for any kind of material and loading condition.

## 4. Discussions

The crack growth direction along the specimen thickness is exceedingly complex in view of the coexistence of shear lips, flat-to-slant zone and the in-plane shear under various out-of-plane constraints. Because crack growth starts in the mid-plane of the specimen where the out-of-plane constraint is highest, only the initiation angle at the mid-plane is taken to validate the analytical predictions. The maximum tangential stress (MTS) criterion [21] and the minimum strain energy density criterion (S-criterion) [22] are the most widely used criterions for predicting crack initiation direction. The governing equation of the MTS-criterion is as follow

$$\theta_0 = \frac{180}{\pi}\left\{4\tan^{-1}\left[\frac{D}{2}+\frac{1}{2}\sqrt{D^2+8}-\frac{1}{2}\frac{\sqrt{2D^2\sqrt{D^2+8}+12\sqrt{D^2+8}+2D^3+16D}}{(D^2+8)^{0.25}}\right]\right\}, \quad (2)$$

where $D = \tan\beta_{eq}$, and $\beta_{eq}$ is the so called loading mixity parameter, or the equivalent crack angle defined by

$$\beta_{eq} = \tan^{-1}\left|\frac{K_I}{K_{II}}\right|. \quad (3)$$

Where $K_I$ and $K_{II}$ are the mode I and mode II stress intensity factors, respectively. For the CTS specimen, the stress intensity factors can be expressed by

$$K_I = \frac{P}{wB}\sin\phi\sqrt{\pi a}\,f_1\left(\phi,\frac{a}{w}\right),$$
$$K_{II} = \frac{P}{wB}\cos\phi\sqrt{\pi a}\,f_2\left(\phi,\frac{a}{w}\right); \quad (4)$$

where $P$ is the applied load, $a$ is the crack length, $w$ is the width of the specimen, $B$ is the thickness of the specimen, and $f_1(\phi,a/w)$ and $f_2(\phi,a/w)$ are geometry factors which can be found in Ref. [33].

The relationship between $\phi$ and $\beta_{eq}$ can be obtained from the above formulas, and the angles used in the present work are listed in Table 2.

The S-criterion can be expressed as

$$S = r\frac{dW}{dV} = r[\frac{1}{2E}(\sigma_x^2+\sigma_y^2+\sigma_z^2)-\frac{v}{E}(\sigma_x\sigma_y+\sigma_y\sigma_z+\sigma_z\sigma_x)+\frac{1}{2\mu}\tau_{xy}^2],$$
$$\frac{\partial S}{\partial \theta}=0, \quad \frac{\partial^2 S}{\partial \theta^2}>0; \quad (5)$$

where, $\mu = \dfrac{E}{2(1+v)}$, $\sigma_x$, $\sigma_y$, $\sigma_z$ and $\sigma_{xy}$ are the stress components.

Another criterion which tries to consider the effect of triaxial stress states on cracking under mixed-mode loading is the maximum stress triaxiality criterion (M-criterion) [23]. It states that the direction of crack initiation should be coincided with the direction of the maximum stress triaxiality (the ratio of the hydrostatic stress to the equivalent stress) along a constant radius around the crack-tip. Mathematically, the M-criterion can be obtained from the elastic solution as

$$tg^4\frac{\theta_0}{2} - 3Dtg^3\frac{\theta_0}{2} - (1-2D^2)tg^2\frac{\theta_0}{2} + \frac{1}{2}(1-D^2)Dtg\frac{\theta_0}{2} - \frac{1+D^2}{2} = 0. \qquad (6)$$

Figure 14 compares the theoretical predictions by the MTS-criterion, S-criterion and M-criterion with the experimental results of crack initiation angle varying with the equivalent crack angle $\beta_{eq}$ for the four thicknesses. It is seen that all the criterions have the same value in pure mode I case, where the crack growth straight forward with zero initiation angle. It is confirmed by the experimental data for all the thickness considered in this work. When mixed-mode I/II loading is applied, however, discrepancy among the criterions, and between the criterions and the experimental data appears. When $\beta_{eq}$ is higher than about 70°, the difference among the three criterions is very small. When $\beta_{eq}$ is less than about 65°, the M-criterion deviates the other two criterions and become too high in comparison with the experimental data. In range of $\beta_{eq}$>30°, the discrepancy between the MTS-criterion and the S-criterion is relatively small, although the S-criterion is higher for smaller $\beta_{eq}$. Plane stress and plane strain states make no difference in the MTS-criterion and only small difference (less than 3°) on $\theta_0$ in the S-criterion. In overall, both the MTS-criterion and the S-criterion provide reasonable predictions of the crack initiation angle for $\beta_{eq} \geq 30^{o}$ ($\phi \geq 15°$) in comparing with the experimental data of the 2, 4 and 14 mm thick specimens except some individual points of large scatter.

In the case of 8 mm thick specimens, however, it is surprising to find that all the experimental data in the mixed-mode range are significantly and systematically lower than all the theoretical predictions. The discrepancy in the crack initiation angle is as high as $15^{o} \sim 20^{o}$ which cannot be explained by uncertainties in the experiment. This phenomenon has not been reported in any material in the literatures for the best knowledge of the authors, so that it cannot be explained yet by the possible change of material structures and properties along the thickness direction of the mother plate. We call it an 8mm-phenomenon for the tested LC4-CS alloy.

We believe that the 8mm-phenomenon is a compositive effect of thickness induced variation in 3D constraint or stress state near the crack front, the ductility and damage mechanism of the materials, the confined plastic deformation and the mixed-mode loading. Of course, temperature and corrosion environments would also have potential influences if they are included. From the view point of 3D constraint, the out-of-plane stress constraint factor $T_z = \sigma_{zz}/(\sigma_{xx} + \sigma_{yy})$ which has most direct relation with thickness is confined mainly within a distance half of the thickness from the crack front [4, 5, 34]. In the interior of a specimen, $T_z$ reaches its upper bound of plane strain value of elastic or elastic-plastic Poisson's ratio of the material at the crack border, and disappear at distance about $B/2$ from the border. Within the region of $r<B/2$, $T_z$ decreases monotonically with increasing distance from the border, and the plane stress state become dominated outside this region. Therefore, in very thin specimens, the 3D transition region is very small and the plane stress dominated region goes very near to the crack border, the 2D theory becomes more efficient, such as in the 2 mm specimen in this study. For very thick specimens, the 3D region becomes large enough and the plane strain constraint becomes dominated in the damaged zone ahead of the crack-tip, so the 2D theory can also be efficient, such as the case of the 14 mm specimens in this study. However, for specimens have thickness in the middle range, transition from plane stress domination to plane strain domination will occur. What makes the things in this transition range out of the two limit bounds of plane stress and plane strain is the complex 3D distribution of $T_z$ near the crack border. It has been generally recognized that ductile damage of materials is closely related to the local 3D stress state. Thus, for certain thickness, the stress state change from point to point within the damage zone which will lead to different ductile damage mechanism and complex damage history in each material element in consideration and the plastic deformation will also become complex near the crack border. Planar approximations or some combinatorial model of plane stress and plane strain are too simple to the complex situation. The 8mm-phenomenon is a typical example of such a transition process.

It must be realized that the transition region of thickness may shift from material to material, mainly associated with the ductility and plasticity of the materials. For example, in very brittle materials the plane strain domination will sustain to extremely thin plate, while in materials with high ductility the plane stress domination can sustain to much large thickness. The fractographic investigations have shown us that the micro mechanism of fracture of the LC4-CS alloy is ductile dimples dominated, but the overall ductility is relatively low. This should be the main reason for the 8mm-phenomenon shown in this study.

Recently, more and more deep investigations show the subtle effects of slight shear stress or deformation on material strength [38, 39]. Such fundamental researches are mainly confined to two-axial loading situations [40], and the complete 3D picture is far from clear up to now. The present experiments just contribute one additional example for the complex picture, and we are unable to give an exact solution to it in the present paper. Further efforts are needed from not only material and mechanical aspects, but also physics level.

For pure mode II crack initiation angle, data are only obtained from the 14 mm and 8 mm specimens. It is about $18^o$~$20^o$, much lower than all the theoretical predictions, but not down to zero as expected by the pure shear fracture theory. As deformation is inevitable, in fact, pure mode II loading condition is hard to be guaranteed in the present CTS tests. For the 4 and 2 mm thick specimens, plastic deformation is too severe in mode II loading to obtain meaningful initiation angle.

5. **Conclusions**

A systematical experimental investigation is performed on the effect of thickness on mixed-mode fracture of the low ductile LC4-CS aluminum alloy with compact-tension-shear specimens of thicknesses ranged from 2 to 14 mm and loading angle ranged from pure mode I to pure mode II in the ambient condition. Loading versus crack opening displacement curves are recorded, in-site crack growth tracing by a CCD camera is adopted. After the test, macro- and micro-fractographic analyses, and careful crack initiation angle measurements at the mid-cutting plane of the specimens are conducted. Comparison with selected theoretical criteria is also made. The following conclusions can be drawn from the study.

(1) The coupled effects of specimen thickness and mixed-mode loading on the mixed-mode loading capacity are significant. Under pure mode I loading condition, the loading capacity per thickness is a monotonically decreasing function of the thickness of specimens. The loading capacity of the 14 mm specimen is only 56% of that of the 2 mm one. However, under the mixed-mode I/II loading condition, the dependence of loading capacity on the thickness becomes very weak, even when the loading angle is only 15$^o$ from the pure mode I direction. The loading capacity of the 4 mm thick specimens is slightly higher than that of the rest specimens in most cases. With the increase of mode II loading fraction, the thickness effect becomes weaker, and almost disappears under pure mode II loading condition.

(2) The measured crack initiation angles from the mid-plane of the broken specimens coincide quite well with the maximum-tangent-stress criterion and the minimum strain energy density criterion in the thin (2 and 4 mm) and thick (14 mm) specimens under all the mixed-mode conditions except pure mode II. However, exceptional low crack initiation angles are found in the 8 mm thick specimens subjected to mixed-mode loading. The discrepancy of the measured crack initiation angles to the predicted values is as high as $15^\circ \sim 20^\circ$ in the mixed-mode cases, although exact coincidence exists in the pure mode I test between them. This strange 8mm-phenomenon should be a combinatorial effect of thickness induced variation in the 3D constraint or stress state near the crack front, the ductility and damage mechanism of the materials, the confined plastic deformation and the mixed-mode loading, but it is requires deeper interdisciplinary study.

(3) The specimen thickness has a strong influence on the macroscopic fracture surface morphology. Except for the pure mode II loading condition, the 2 mm and 4 mm thick specimens show slant macroscopic fracture surfaces; the 14 mm thick specimens show flat fracture in the center of the plate with shear lips near the free surfaces of the specimens; the 8 mm thick specimens have more complex fractographic appearance comprising flat fracture with shear lips and slant fracture. Thickness effects exist in the microscopic fracture profiles as well. Whatever the specimen thickness is, the characters of ductile failure are distinct due to the void nucleation, growth and coalescence. However, the size of the final voids varies significantly with specimen thickness and loading mode. The ductile dimples in the crack initiation zone become larger and shallower as the specimen thickness decreasing.

**ACKNOWLEDGEMENT**

The NSF of China (No.50275073), the Aeronautic Science Foundation of China (No. 03B52011) and the Cheung Kong Scholars Program supported this work. The authors would like to thank Mr. Yuequan Zhou employing in Northwestern Polytechnic University for his invaluable experimental assistance.

**Table Captions List**

Table 1 Chemical composition of the material (wt %)

Table 2 The equivalent crack angle $\beta_{eq}$ versus the loading angle $\phi$ (degree)

Table 1 Chemical composition of the material (wt %)

| Mg | Cu | Zn | Al |
|---|---|---|---|
| 4.28 | 2.50 | 6.51 | Remain |

Table 2 The equivalent crack angle $\beta_{eq}$ versus the loading angle $\phi$ (degree)

| $\phi$ | 90 | 75 | 60 | 45 | 30 | 15 | 0 |
|---|---|---|---|---|---|---|---|
| $\beta_{eq}$ | 90 | 83 | 75 | 65 | 51 | 31 | 0 |

**Figures List**

Fig. 1 Shape and size of the compact-tension-shear (CTS)

Fig. 2 Loading frame and definition of the loading angle

Fig. 3 Crack tip appearance during loading process. The major tic increment is 1 mm in the ruler.

Fig. 4 Load-COD curves for different thicknesses under loading conditions of (a) pure mode Ⅰ, =90°, and mixed-mode with (b) =60°, (c) =45°, (d) =30° and (e) =15°

Fig. 5 The maximum load per unit thickness against specimen thickness under pure mode II loading.

Fig. 6 The effective stress intensity factor with initial crack length versus COD curves

Fig. 7 Load-COD curves for different loading angles of specimens with thickness of (a) 14 mm, (b) 8 mm, (c) 4 mm and (d) 2 mm.

Fig. 8 (a) Schematic typical fracture surface given by **Mahmoud and Lease (2003); James and Newman Jr (2003)**. (b) Schematic typical fracture surface for the 14 mm specimens under mixed-mode loading. (c). Schematic typical fracture surface for the 2 and 4 mm specimens under mixed-mode loading.

Fig. 9 The SEM fractographs in the crack initiation region under mode I loading condition for specimens of (a) B=14 mm, (b) B=8 mm, (c) B=4 mm.

Fig. 10 The SEM fractographs in the crack initiation region under mode Ⅱ loading condition for specimens of (a) B=14 mm, (b) B=8 mm.

Fig. 11 The SEM fractographs in the crack initiation region under 45° loading

condition for specimens of (a) B=14 mm, (b) B=8 mm, (c) B=4 mm, (d) B=2 mm, (e) B=2 mm.

Fig. 12 Definition and measurement of crack initiation angle

Fig. 13 Experimental results of crack initiation angle versus thickness in dependence on the loading angle

Fig. 14 Crack initiation angle $\theta_0$ versus $\beta_{eq}$ curves for specimens of (a) B=14 mm, (b) B=8 mm, (c) B=4 mm, (d) B=2 mm.

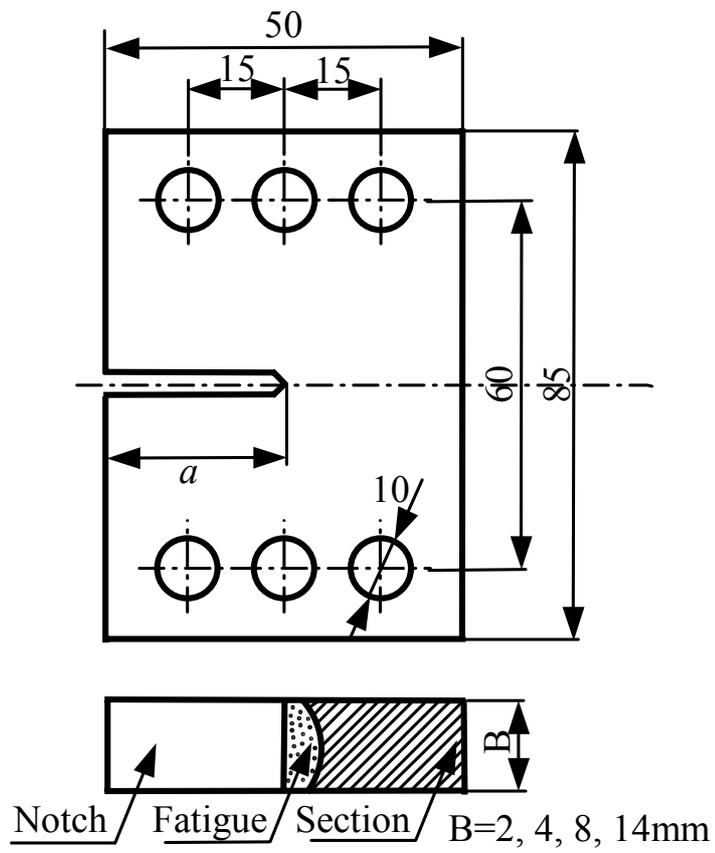

Fig. 1 Shape and size of the compact-tension-shear (CTS)

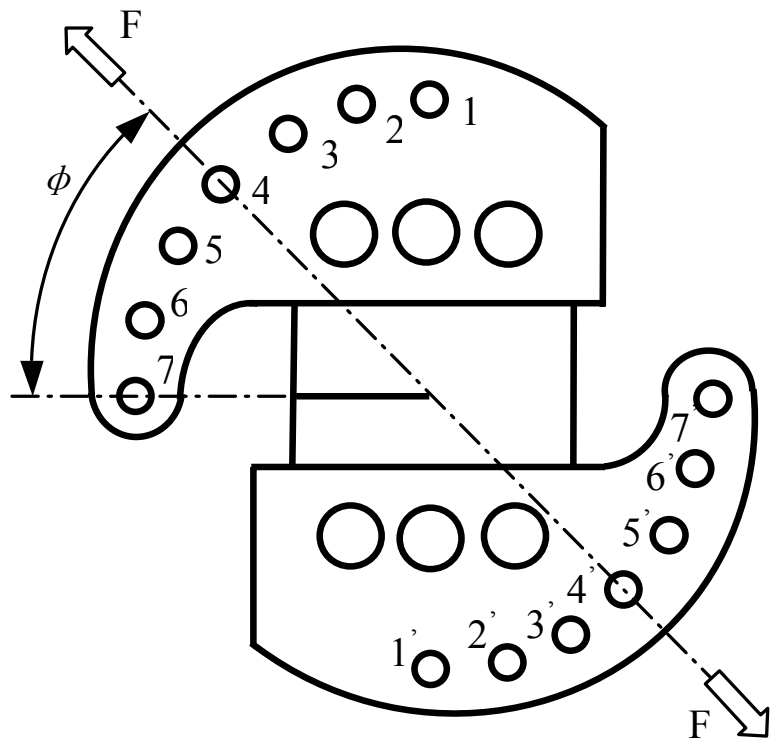

Fig. 2 Loading frame and definition of the loading angle

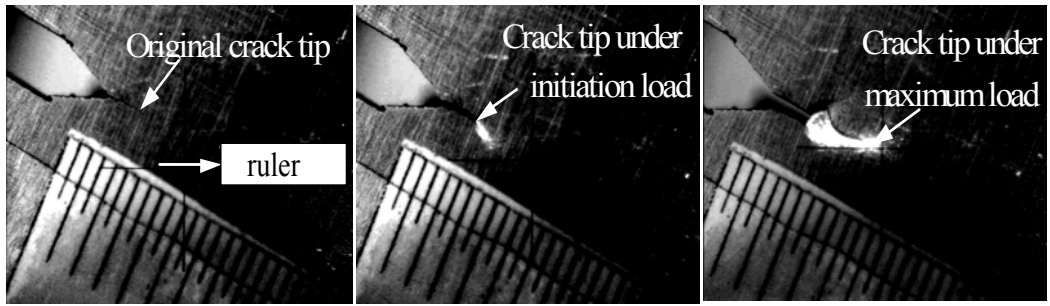

(a) initial crack at $a_0$   (b) crack at initiation point $a_i$   (c) crack at maximum load

Fig. 3 Crack tip appearance during loading process. The major tic increment is 1 mm in the ruler.

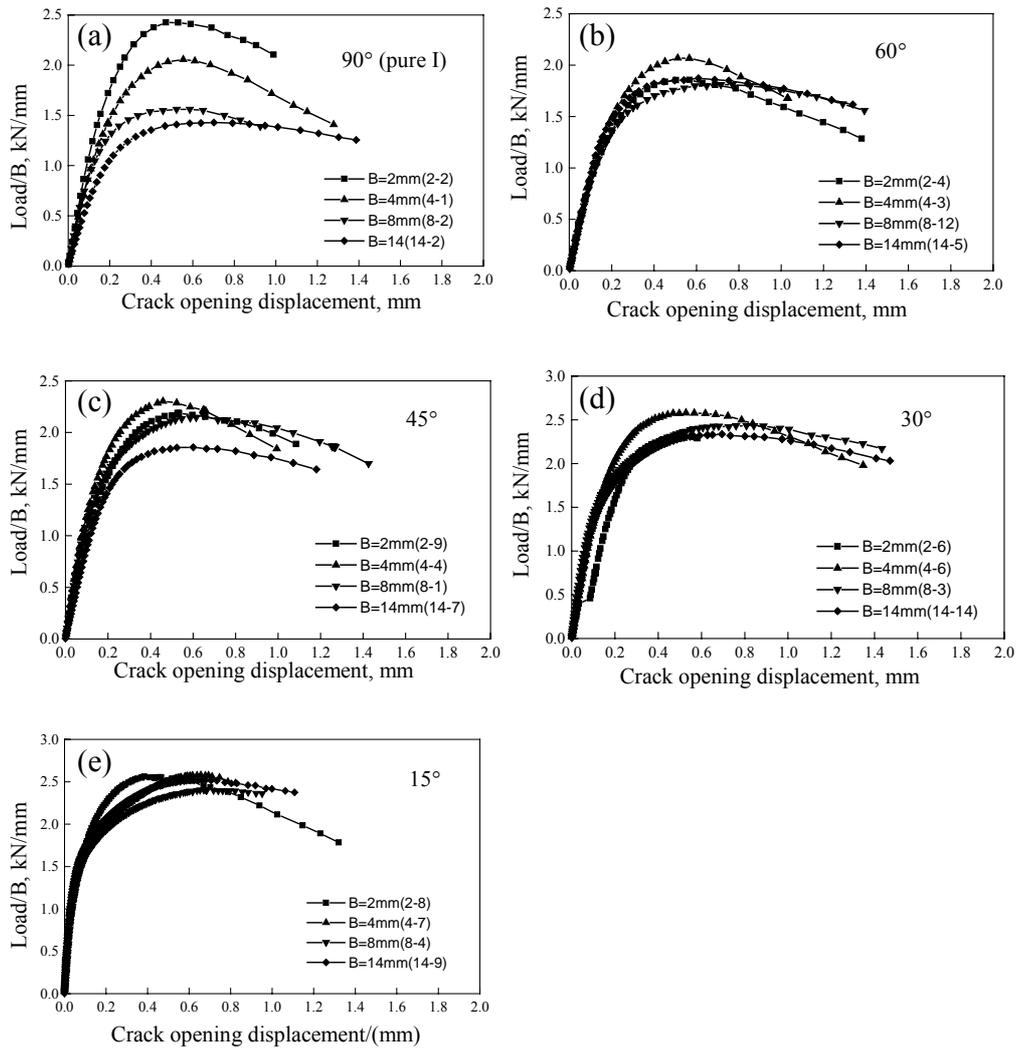

Fig. 4 Load-COD curves for different thicknesses under loading conditions of (a) pure mode Ⅰ, =90°, and mixed-mode with (b) =60°, (c) =45°, (d) =30° and (e) =15°.

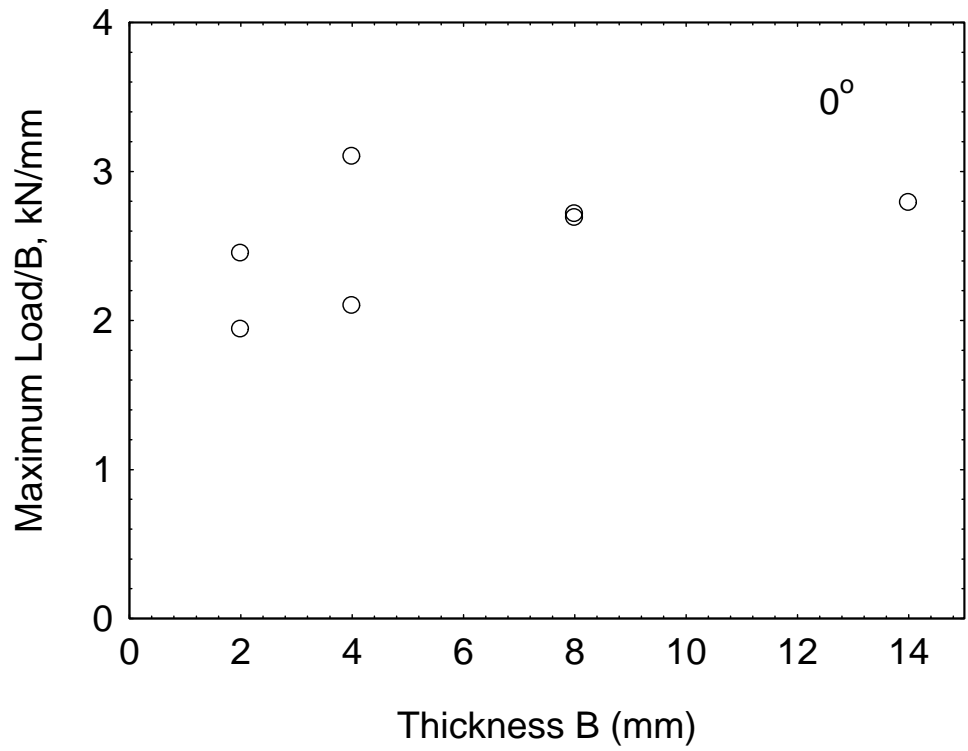

Fig. 5 The maximum load per unit thickness against specimen thickness under pure mode II loading.

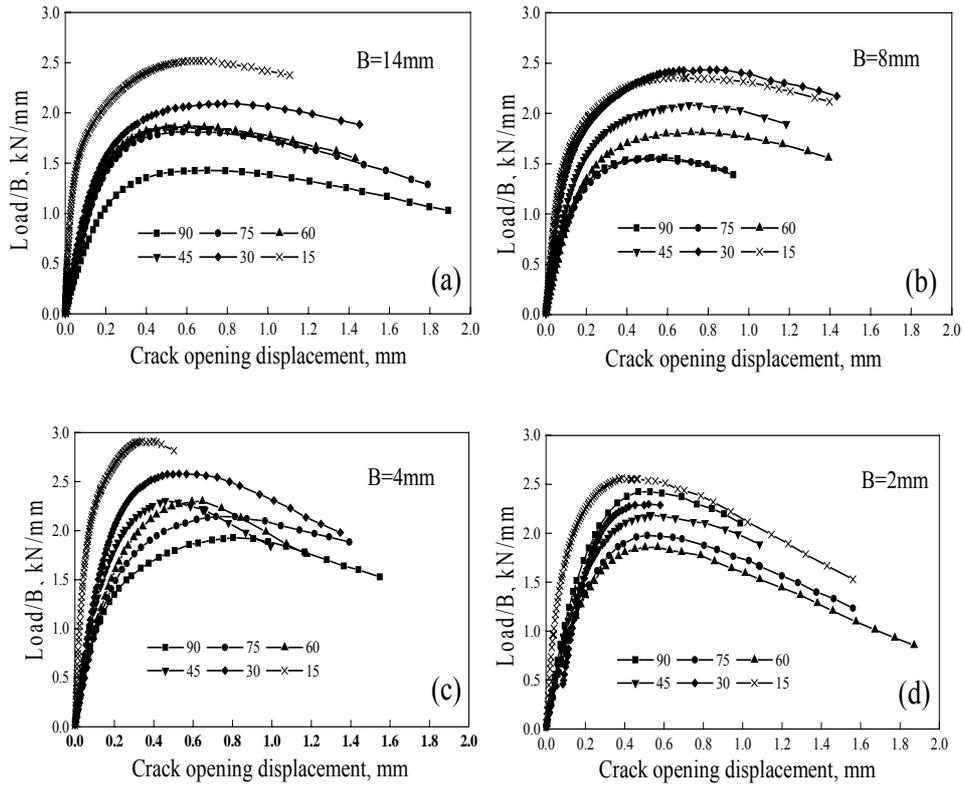

Fig. 6 The effective stress intensity factor with initial crack length versus COD curves

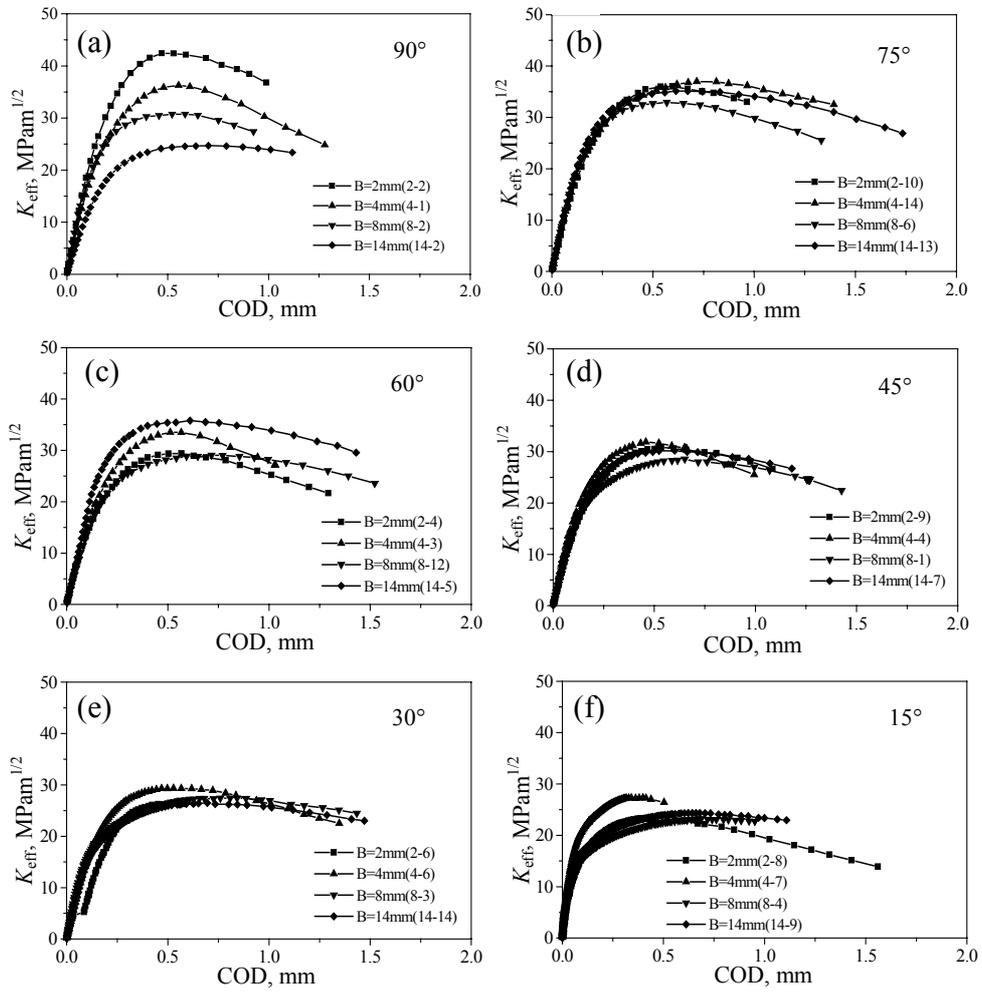

Fig. 7 Load-COD curves for different loading angles of specimens with thickness of (a) 14 mm, (b) 8 mm, (c) 4 mm and (d) 2 mm.

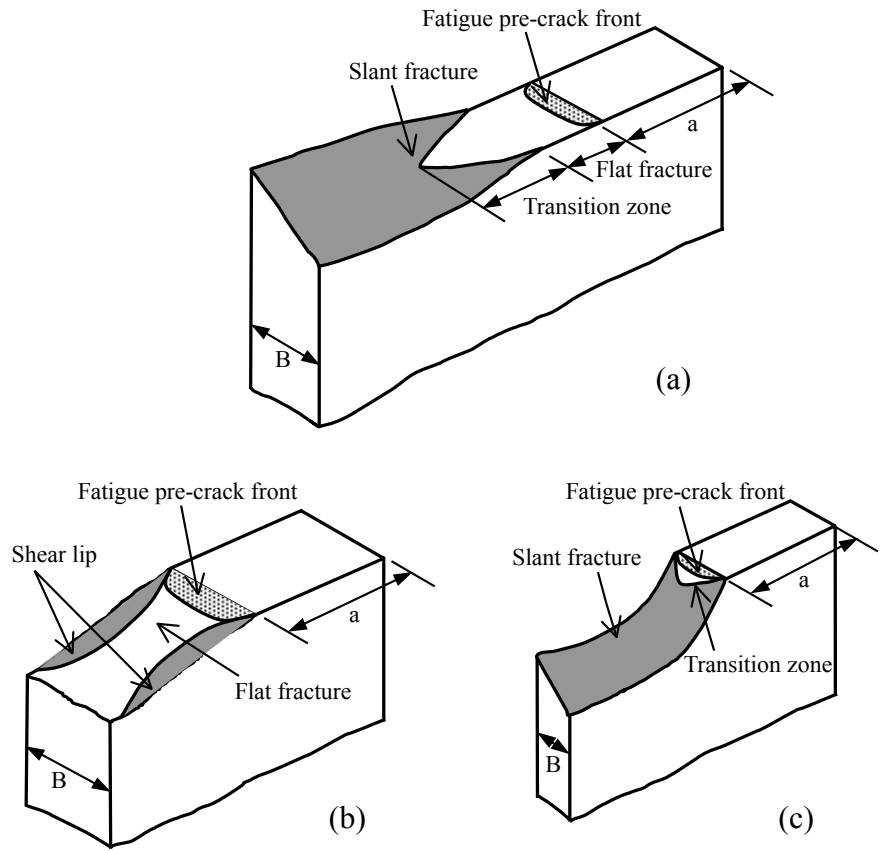

Fig. 8. (a) Schematic typical fracture surface given by **Mahmoud and Lease (2003); James and Newman Jr (2003)**. (b) Schematic typical fracture surface for the 14 mm specimens under mixed-mode loading. (c). Schematic typical fracture surface for the 2 and 4 mm specimens under mixed-mode loading.

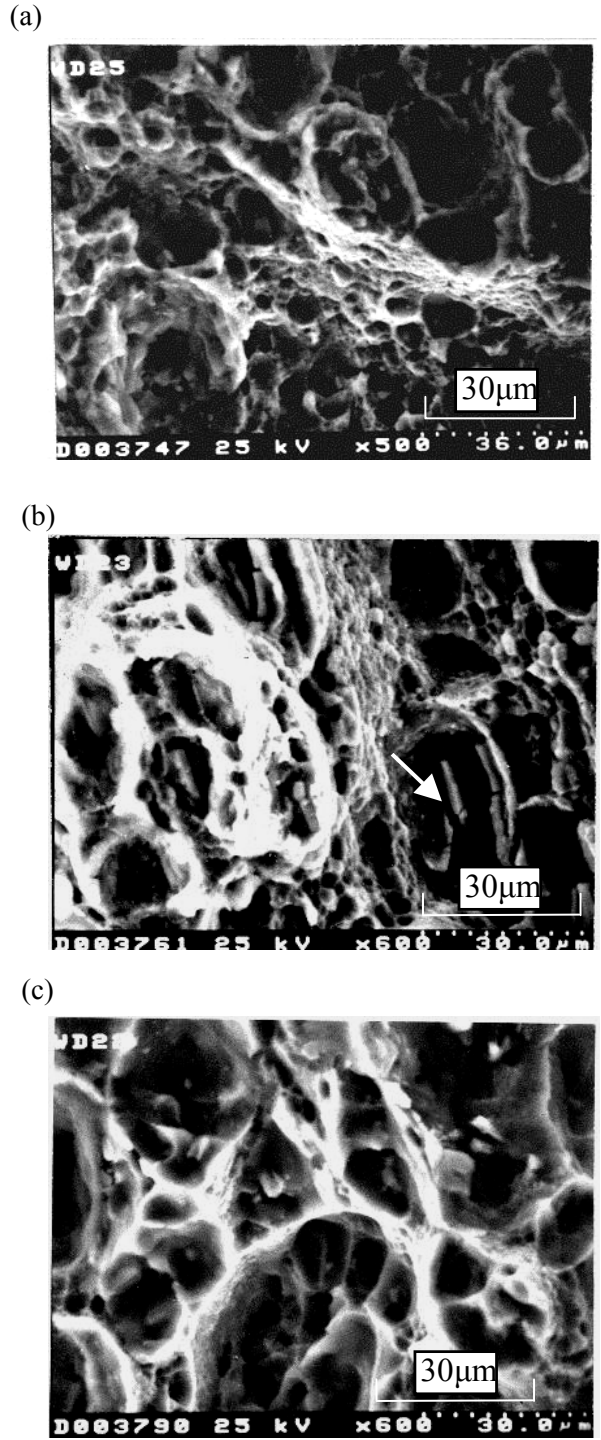

Fig. 9 The SEM fractographs in the crack initiation region under mode I loading condition for specimens of (a) B=14 mm, (b) B=8 mm, (c) B=4 mm.

(a)

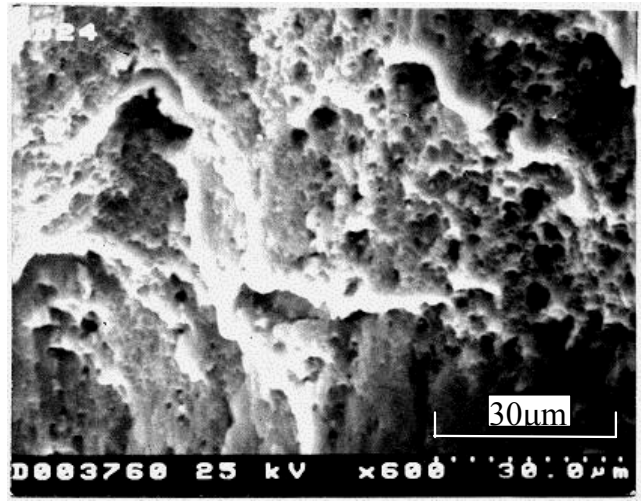

(b)

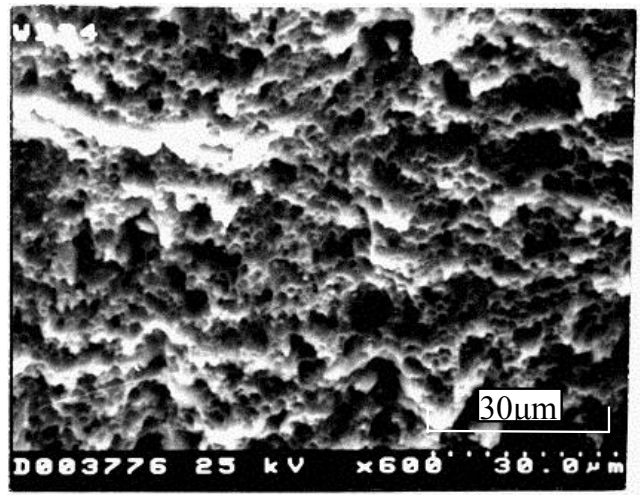

Fig. 10 The SEM fractographs in the crack initiation region under mode II loading condition for specimens of (a) B=14 mm, (b) B=8 mm.

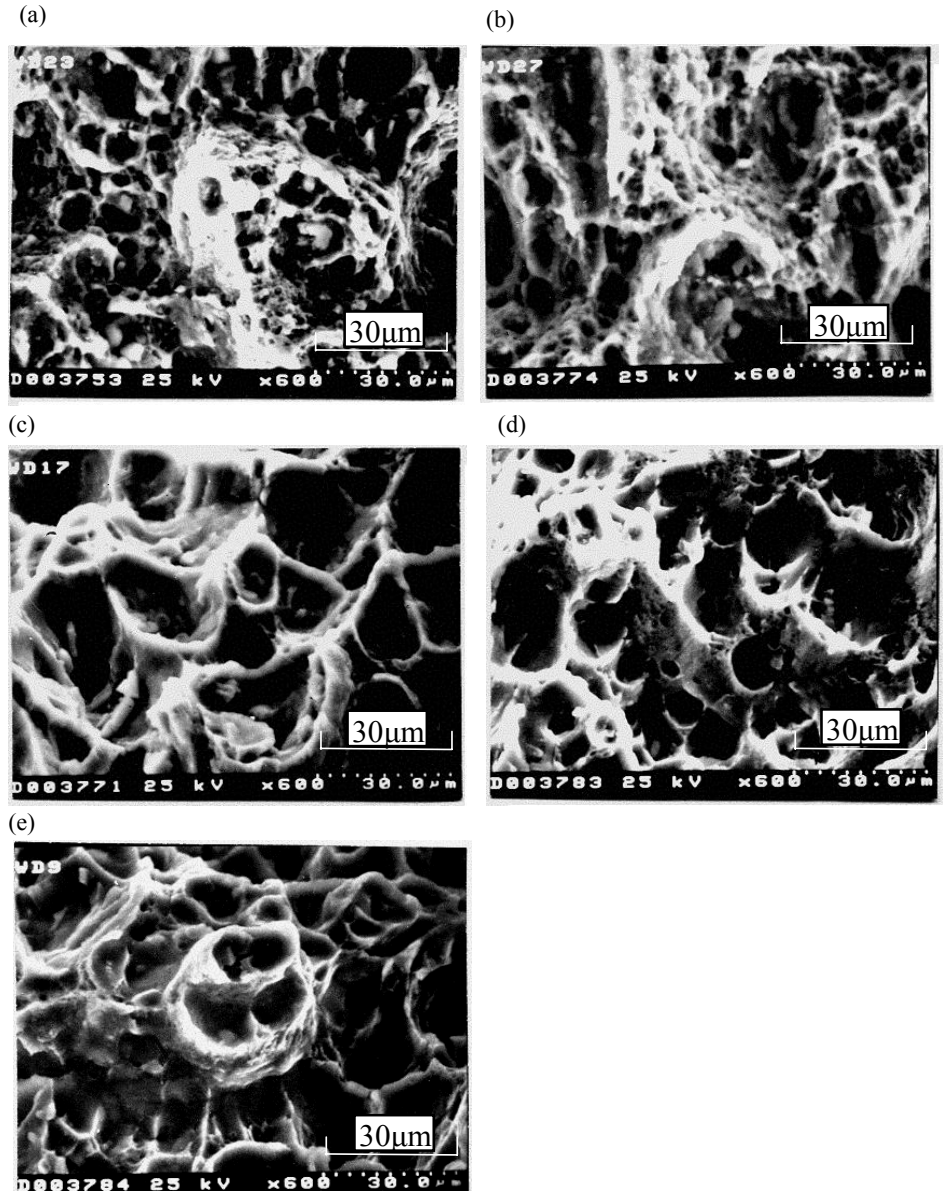

Fig. 11 The SEM fractographs in the crack initiation region under 45° loading condition for specimens of (a) B=14 mm, (b) B=8 mm, (c) B=4 mm, (d) B=2 mm, (e) B=2 mm.

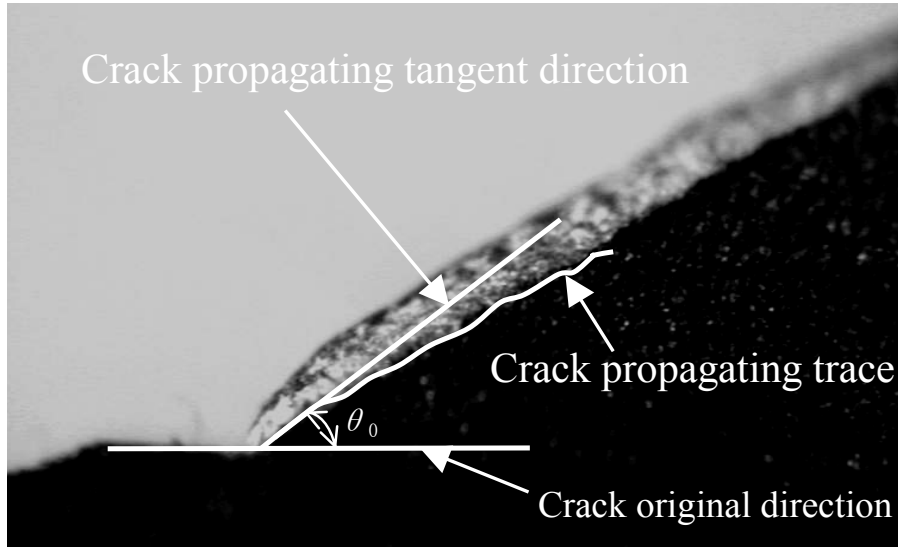

Fig. 12 Definition and measurement of crack initiation angle

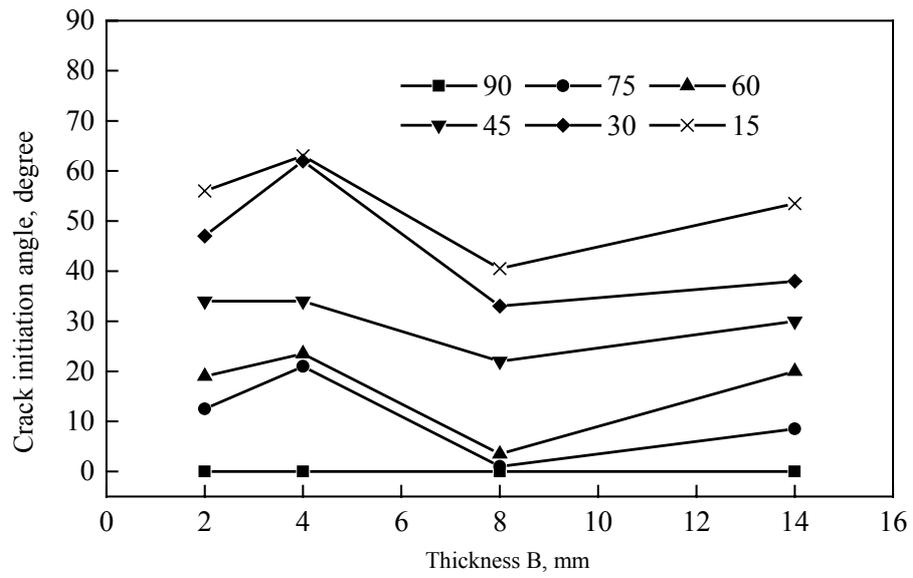

Fig. 13 Experimental results of crack initiation angle versus thickness in dependence on the loading angle

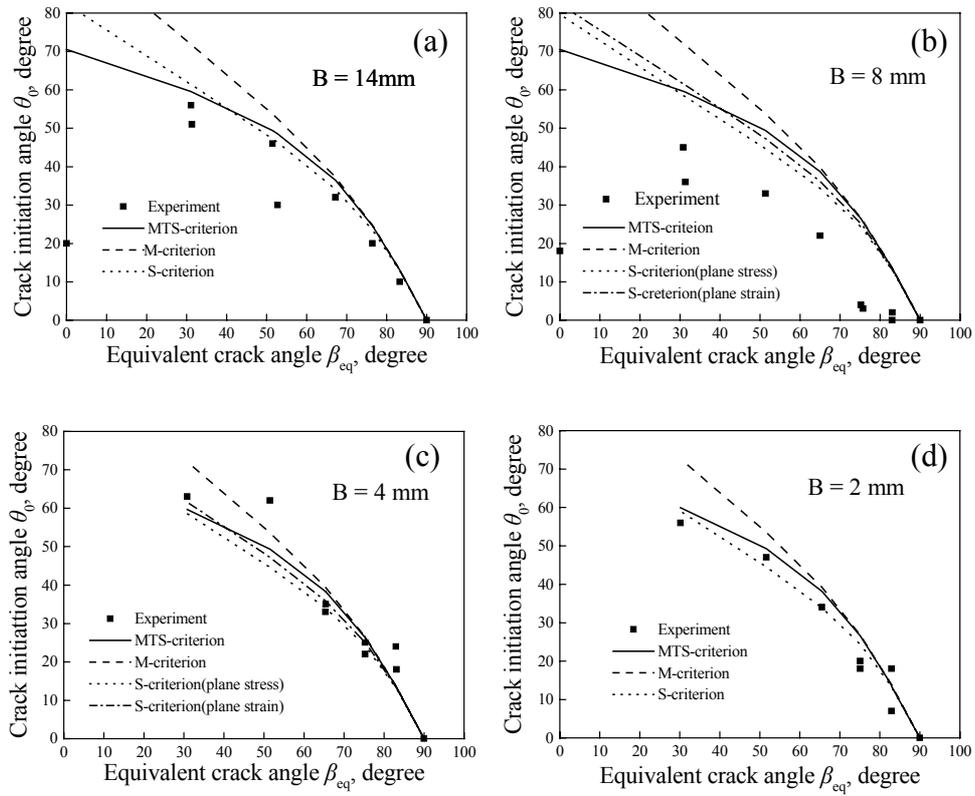

Fig. 14 Crack initiation angles $\theta_0$ versus $\beta_{eq}$ curves for specimens of (a) B=14 mm, (b) B=8 mm, (c) B=4 mm, (d) B=2 mm.